\documentclass[doublecol]{epl2}

\title{Parameters of the fractional Fokker-Planck equation}

\author{S. I. Denisov\inst{1,2} \and Peter H\"{a}nggi\inst{3}
\and Holger Kantz\inst{1}} \shortauthor{S. I. Denisov \etal}

\institute{
  \inst{1} Max-Planck-Institut f\"{u}r Physik komplexer Systeme -
  N\"{o}thnitzer Stra{\ss}e 38, D-01187 Dresden, Germany\\
  \inst{2} Department of Physics, Sumy State University -
  2 Rimsky-Korsakov Street, 40007 Sumy, Ukraine\\
  \inst{3} Institut f\"{u}r Physik, Universit\"{a}t Augsburg -
  Universit\"{a}tsstra{\ss}e 1, D-86135 Augsburg, Germany

  }

\abstract{ We study the connection between the parameters of the fractional
Fokker-Planck equation, which is associated with the overdamped Langevin
equation driven by noise with heavy-tailed increments, and the transition
probability density of the noise generating process. Explicit expressions for
these parameters are derived both for finite and infinite variance of the
rescaled transition probability density.}

\begin{document}

\maketitle

\section{Introduction}
Heavy-tailed distributions, i.e., pro\-bability distributions with power tails
and infinite second moments, are an important tool for studying a number of
physical, biological, economical and other systems whose behavior is determined
by rare but large events \cite{SZF,MK, BMR,Rach}. In many cases the
continuous-time dynamics of these systems can be effectively described by the
(dimensionless) overdamped Langevin equation
\begin{equation}
    \dot{x}(t) = f(x(t),t) + \xi(t),
\label{Langevin}
\end{equation}
where $x(t)$ [$x(0)=0$] is a state parameter of the system, $f(x,t)$ is a
deterministic function, and $\xi(t)$ is a random noise defined by the
infinitesimal increments $\Delta\eta(t) = \int_{t} ^{t + \tau} dt' \xi(t')$
($\tau \to 0$) that are assumed to be independent on non-overlapping intervals
and distributed with a heavy-tailed distribution. Since the tails of these
distributions can not be neglected, the classical stochastic theory, which is
based on the ordinary central limit theorem, is not applicable to
eq.~(\ref{Langevin}). Specifically, if the increments are distributed according
to a L\'{e}vy stable distribution~\cite{Zol}, i.e., $\xi(t)$ is a L\'{e}vy
stable noise, then the probability density $P(x,t)$ that $x(t)=x$ satisfies the
\textit{fractional} Fokker-Planck equation~\cite{JMF,Dit, YCST,BS, EK,DS} which
can by written as
\begin{eqnarray}
    \frac{\partial}{\partial t}P(x,t) \!\!\!&=&\!\!\! -\frac{\partial}
    {\partial x}f(x,t)P(x,t) + \gamma \frac{\partial^{\alpha}}
    {\partial |x|^{\alpha}}\,P(x,t)
    \nonumber\\[6pt]
    &&\!\!\! + \, \gamma \beta \tan\frac{\pi\alpha}{2}\,\frac{\partial}
    {\partial x}\, \frac{\partial^{\alpha -1}}
    {\partial |x|^{\alpha-1}}\,P(x,t).
    \label{FP1}
\end{eqnarray}
Here, the Riesz derivative, $\partial^{\alpha}/\partial |x|^{\alpha}$, is
defined as~\cite{SKM} $\partial^{\alpha}h(x)/\partial |x|^{\alpha} =
-\mathcal{F}^{-1} \{ |k|^{\alpha} h_{k} \}$, a pair $\mathcal{F} \{ h(x) \}
\equiv h_{k} = \int_{-\infty}^{\infty} dx\,e^{-ikx}h(x)$ and $\mathcal{F}^{-1}
\{ h_{k} \} \equiv h(x) = (1/2\pi)\times \int_{-\infty}^{\infty} dk\,e^{ikx}
h_{k}$ represents the Fourier transforms, and $\alpha$, $\beta$ and $\gamma$
are the parameters of the stable distribution.

Because of the generalized central limit theorem \cite{GK}, the Levy stable
distributions constitute an important but a particular class of heavy-tailed
distributions. In this Letter, we show that the fractional Fokker-Planck
equation~(\ref{FP1}) is valid also for \textit{all} noises $\xi(t)$ whose
increments have heavy-tailed distributions. Explicit expressions for the
parameters of eq.~(\ref{FP1}) are derived in terms of the asymptotic
characteristics of these distributions.

\section{Definitions and basic equations}
Our starting point is the generalized Fokker-Planck equation~\cite{DHH1}
\begin{equation}
    \frac{\partial}{\partial t}P(x,t) = -\frac{\partial}{\partial x}
    f(x,t)P(x,t) + \mathcal{F}^{-1}\{P_{k}(t) \ln S_{k}\},
    \label{FP2}
\end{equation}
which corresponds to the Langevin equation (\ref{Langevin}) driven by an
\textit{arbitrary} noise. The term ``arbitrary" means that the independent
increments $\Delta\eta(j\tau) = \eta(j\tau + \tau) - \eta(j\tau) = \int_{j\tau}
^{j\tau + \tau} dt' \xi(t')$ ($\tau \to 0$, $j=0,1,\ldots$) of the
discrete-time noise generating process $\eta(n\tau) = \sum_{j=0}^{n-1} \Delta
\eta(j\tau)$ ($n = 1,2,\ldots$) are distributed according to an arbitrary
probability density function $p(\Delta\eta, \tau)$. In other words,
$p(\Delta\eta, \tau)$ is the transition probability density of the process
$\eta(n\tau)$. It is assumed that (i) $p(\Delta\eta, \tau)$ is properly
normalized, i.e., $\int_{-\infty}^ {\infty} d(\Delta \eta)\, p(\Delta
\eta,\tau) = 1$, (ii) the first moment, if it exists, equals zero, i.e.,
$\int_{-\infty}^{\infty} d(\Delta \eta)\, p(\Delta \eta,\tau) \Delta \eta = 0$,
and (iii) $\lim_{\tau \to 0} p(\Delta \eta,\tau) = \delta(\Delta \eta)$, where
$\delta(\cdot)$ is the Dirac $\delta$ function. The characteristic function
$S_{k} = \langle e^{-ik\eta(1)} \rangle$ of $\eta(1) = \lim_{\tau \to 0}
\sum_{j=0}^{[1/\tau]-1} \Delta \eta (j\tau)$ ($[1/\tau]$ denotes the integer
part of $1/\tau$) is connected with the characteristic function $p_{k} (\tau) =
\langle e^{-ik\Delta \eta(j\tau)} \rangle$ of $\Delta \eta(j\tau)$, i.e., the
Fourier transform of $p(\Delta\eta, \tau)$, via the relation~\cite{DHH1}
\begin{equation}
    \ln S_{k} = \lim_{\tau \to 0} \frac{1}{\tau}[p_{k}(\tau) - 1].
    \label{S}
\end{equation}

The generalized central limit theorem~\cite{GK} implies that for a wide class
of properly scaled probability densities $p(\Delta\eta, \tau)$ the
characteristic function $S_{k}$ corresponds to L\'{e}vy stable distributions.
These distributions are described by four parameters~\cite{Zol}: an index of
stability $\alpha \in (0,2]$, a skewness parameter $\beta \in [-1,1]$, a scale
parameter $\gamma \in (0, \infty)$, and a location parameter $\rho \in
(-\infty, \infty)$, which, in accordance with the initial condition $P(x,0) =
\delta(x)$, is assumed to be zero. Therefore, excluding from consideration the
singular case when $\alpha = 1$ and $\beta \neq 0$ simultaneously (in this case
$|\ln S_{k}| = \infty$ if $k \neq 0$, see below, and as a consequence the
system reaches the final state immediately), one obtains
\begin{equation}
    S_{k} = \exp\left[ -\gamma |k|^{\alpha} \left(1 + i\beta\,
    \rm{sgn}(\it{k}) \tan \frac{\pi\alpha}{\rm 2}\right)\right],
    \label{charfunct}
\end{equation}
and eq.~(\ref{FP2}) reduces to eq.~(\ref{FP1}) \cite{DHH1}.

In order to find the stable parameters in eq.~(\ref{FP1}), the transition
probability density $p(\Delta\eta, \tau)$ must be specified. Next we consider a
class of the functions $p(\Delta\eta, \tau)$ defined as
\begin{equation}
    p(\Delta\eta,\tau) = \frac{1}{a(\tau)}p\left( \frac{\Delta\eta}
    {a(\tau)} \right).
    \label{TPD}
\end{equation}
Here, $a(\tau)$ is a positive scale function that vanishes at $\tau = 0$, and
the rescaled transition probability density, $p(y)$, is an arbitrary
probability density which satisfies the condition $\lim_ {\epsilon \to 0}
p(y/\epsilon)/ \epsilon = \delta (y)$ and has zero first moment (if it exists).
According to this definition, $p_{k} (\tau) = p_{ka(\tau)}$ and the
normalization condition for $p(y)$, which is equivalent to $p_{0}=1$, yields
$\ln S_{0} = 0$. If $k\neq0$, then $|\ln S_{k}| \in [0,\infty]$ and we can
select three physically different situations depending on how quickly $a(\tau)$
tends to zero as $\tau \to 0$. First, if $p_{ka(\tau)} - 1 = o(\tau)$ (the
scale function quickly vanishes), then $\ln S_{k} = 0$ and the noise is so weak
that it does not effect the system at all. Second, if $p_{ka(\tau)} - 1$ tends
to zero slower than $\tau$ (the scale function slowly vanishes), then $|\ln
S_{k}| = \infty$, i.e., the influence of the noise is so strong that the system
relaxes instantaneously to the final state. Finally, the case we are primarily
interested in corresponds to $p_{ka(\tau)} - 1 = O(\tau)$, i.e., $0 < |\ln
S_{k}| < \infty$ and the noise acts on the system in a non-trivial way.

Using eqs.~(\ref{S}), (\ref{TPD}) and the representation $\ln S_{k} = R(k) +
iI(k)$, we find the real,
\begin{equation}
    R(k) = -\lim_{\tau \to 0} \frac{1}{\tau} \int_{-\infty}^{\infty}
    dy\, p(y)[1 - \cos(ka(\tau)y)],
    \label{R}
\end{equation}
and imaginary,
\begin{equation}
    I(k) = -\lim_{\tau \to 0} \frac{1}{\tau} \int_{-\infty}^{\infty}
    dy\, p(y)\sin(ka(\tau)y),
    \label{I}
\end{equation}
parts of $\ln S_{k}$. In the next two sections we will evaluate $R(k)$ and
$I(k)$ and express the stable parameters through a few main characteristics of
$p(y)$ and $a(\tau)$. It should be noted in this context that the results are
quite different for the cases with finite and infinite variance $\sigma^{2} =
\int_{-\infty}^{\infty} dy\, p(y)y^{2}$ of the probability density $p(y)$.
Mathematically, this difference arises from the fact that in the latter case
the operations of taking the limit and integration in eqs.~(\ref{R}) and
(\ref{I}) do not commute.

\section{Density functions with finite variance} If the variance $\sigma^{2}$
is finite then, taking first the limit and then integrating, from eq.~(\ref{R})
we obtain
\begin{equation}
    R(k) = -\frac{k^{2}\sigma^{2}}{2} \lim_{\tau \to 0}
    \frac{a^{2}(\tau)}{\tau}.
    \label{R2}
\end{equation}
Since, by assumption, the first moment of $p(y)$ equals zero, for calculating
$I(k)$ we temporarily assume that the third moment, $m_{3} = \int_{-\infty}^
{\infty}dy\, p(y)y^{3}$, exists. This yields
\begin{equation}
    I(k) = -\frac{k^{3}m_{3}}{6} \lim_{\tau \to 0}
    \frac{a^{3}(\tau)}{\tau}.
    \label{I2}
\end{equation}
As is seen from eqs.~(\ref{R2}) and (\ref{I2}), the condition $0<|\ln S_{k}|
<\infty$ holds only if $a^{2}(\tau) \sim q\tau$ ($0<q<\infty$). In this case
$I(k) = 0$, $\mathcal{F}^{-1}\{P_{k}(t) \ln S_{k}\} = (\sigma^{2}q/2)
\partial^{2} P(x,t) /\partial x^{2}$, and the generalized Fokker-Planck
equation (\ref{FP2}) reduces to the ordinary Fokker-Planck equation \cite{Risk}
\begin{equation}
    \frac{\partial}{\partial t}P(x,t) = -\frac{\partial}
    {\partial x}f(x,t)P(x,t) + \gamma \frac{\partial^{2}}
    {\partial x^{2}}\,P(x,t),
    \label{FP3}
\end{equation}
which has the form of eq.~(\ref{FP1}) with
\begin{equation}
    \alpha = 2, \quad \gamma = \frac{\sigma^{2} q}{2}.
    \label{ag1}
\end{equation}

If the third moment $m_{3}$ does not exist then $I(k)$ can be evaluated by the
method described in the next section, yielding the same result: $I(k)=0$. Thus,
if the increments of the noise generating process have \textit{finite}
variance, the Langevin equation (\ref{Langevin}) is always associated with the
\textit{ordinary} Fokker-Planck equation (\ref{FP3}). In this case the noise
$\xi(t)$ is white, i.e., it has a constant power spectral density at all
frequencies, and $\gamma$ is the white noise intensity. In particular, for
Gaussian white noise of intensity $D$ characterized by the Gaussian probability
density $p(y) = (4\pi D)^{-1/2} \exp[-y^{2}/(4D)]$ and the scale function
$a(\tau) = \tau^{1/2}$, we obtain $\sigma^{2} = 2D$, $q=1$, and so $\gamma =
D$.

\section{Density functions with infinite variance} For noises with $\sigma^{2}
= \infty$ the power spectral density does not exist. To find $R(k)$ and $I(k)$
in this case, we consider a class of probability densities $p(y)$ whose
asymptotic behavior is characterized by heavy tails, i.e.,
\begin{equation}
    p(y) \sim \frac{h_{\pm}}{|y|^{1+ \Lambda_{\pm}}} \quad
    (y \to \pm \infty),
    \label{asymp}
\end{equation}
with $\Lambda = \min\{ \Lambda_{+}, \Lambda _{-}\} \in (0,2]$ and $h_{\pm}>0$.
Let us first calculate $R(k)$ at $\Lambda \in (0,2)$. In this case,
representing eq.~(\ref{R}) in the form
\begin{eqnarray}
    R(k) \!\!\!&=&\!\!\! -\lim_{\tau \to 0} \frac{1}{\tau a(\tau)|k|}
    \int_{0}^{\infty} dz\left[ p\left( \frac{z}{a(\tau)|k|}
    \right) \right.
    \nonumber\\[6pt]
    &&\!\!\! + \left.  p\left( \frac{-z}{a(\tau)|k|} \right) \right]
    (1 - \cos z),
    \label{R3}
\end{eqnarray}
replacing $p(y)$ by the asymptotic formula (\ref{asymp}), and using the
integral relation
\begin{equation}
    \int_{0}^{\infty} dz \frac{1-\cos z} {z^{1+\Lambda}} =
    \frac{\pi}{2\Gamma(1+\Lambda)\sin(\pi\Lambda/2)}
    \label{int1}
\end{equation}
($\Lambda \in (0,2)$, $\Gamma(\cdot)$ is the gamma function), we find
\begin{equation}
    R(k) = - |k|^{\Lambda} \frac{\pi (h_{+}\delta_{\Lambda
    \Lambda_{+}} + h_{-}\delta_{\Lambda \Lambda_{-}})}{2\Gamma(1+
    \Lambda)\sin(\pi\Lambda/2)}\lim_{\tau \to 0} \frac{a^{\Lambda}
    (\tau)}{\tau},
    \label{R4}
\end{equation}
where $\delta_{\Lambda \Lambda_{\pm}}$ is the Kronecker symbol.

At $\Lambda = 2$ the integral in eq.~(\ref{int1}) does not exist. Therefore,
for calculating $R(k)$ in this case, we present the integral in eq.~(\ref{R3})
as a sum of two integrals over the intervals $(0,a(\tau)|k|\xi)$ and
$(a(\tau)|k|\xi, \infty)$ with $\xi = O(1)$. This yields $R(k) = R_{1}(k,\xi) +
R_{2}(k,\xi)$, where
\begin{equation}
    R_{1}(k,\xi) = - \frac{k^{2}}{2} \int_{0}^{\xi} dy\, [p(y) +
    p(-y)]y^{2} \lim_{\tau \to 0} \frac{a^{2}(\tau)}{\tau}
    \label{R(1)}
\end{equation}
and
\begin{eqnarray}
    R_{2}(k,\xi) \!\!\!&=&\!\!\! - k^{2} (h_{+}\delta_{2 \Lambda_{+}}
    + h_{-}\delta_{2 \Lambda_{-}}) \lim_{\tau \to 0} \frac{a^{2}
    (\tau)}{\tau}
    \nonumber\\[6pt]
    &&\!\!\! \times \int_{a(\tau)|k|\xi}^{\infty}dz \frac{1-\cos z}
    {z^{3}}.
    \label{R(2)}
\end{eqnarray}
Since $\int_{a(\tau)|k|\xi}^{\infty}dz\, (1-\cos z) z^{-3} \sim (1/2) \ln
[1/a(\tau)] \to \infty$ as $\tau \to 0$, the first term, $R_{1}(k,\xi)$, can be
neglected in comparison with the second, $R_{2}(k,\xi)$, yielding
\begin{equation}
    R(k) = - k^{2} \frac{h_{+}\delta_{2 \Lambda_{+}} + h_{-}
    \delta_{2 \Lambda_{-}}}{2} \lim_{\tau \to 0} \frac{a^{2}
    (\tau)}{\tau}\ln \frac{1}{a(\tau)}.
    \label{R5}
\end{equation}

In order to find explicit expressions for the imaginary part of $\ln S_{k}$, we
first rewrite eq.~(\ref{I}) in the form
\begin{eqnarray}
    I(k) \!\!\!&=&\!\!\! -\lim_{\tau \to 0} \frac{1}{\tau a(\tau)k}
    \int_{0}^{\infty} dz\left[ p\left( \frac{z}{a(\tau)|k|}
    \right) \right.
    \nonumber\\[6pt]
    &&\!\!\! - \left.  p\left( \frac{-z}{a(\tau)|k|} \right) \right]
    \sin z.
    \label{I3}
\end{eqnarray}
Then, assuming that $\Lambda \in (0,1)$, we substitute eq.~(\ref{asymp}) into
eq.~(\ref{I3}). Finally, taking into account that
\begin{equation}
    \int_{0}^{\infty} dz \frac{\sin z} {z^{1+\Lambda}} =
    \frac{\pi}{2\Gamma(1+\Lambda)\cos(\pi\Lambda/2)}
    \label{int2}
\end{equation}
if $\Lambda \in (0,1)$, eq.~(\ref{I3}) can be reduced to
\begin{equation}
    I(k) = - k|k|^{\Lambda -1} \frac{\pi (h_{+}\delta_{\Lambda
    \Lambda_{+}} - h_{-}\delta_{\Lambda \Lambda_{-}})}{2\Gamma(1+
    \Lambda)\cos(\pi\Lambda/2)} \lim_{\tau \to 0} \frac{a^{\Lambda}
    (\tau)}{\tau}.
    \label{I4}
\end{equation}

If $\Lambda \in (1,2]$ then the integral in eq.~(\ref{int2}) diverges at the
lower limit of integration, and the described approach becomes inapplicable. To
generalize it to $\Lambda \in (1,2]$, we use the condition $\int_{-\infty}^
{\infty} dy\, p(y)y = 0$, which permits us to replace $\sin z$ by $\sin z - z$
in eq.~(\ref{I3}). As a consequence, we arrive to the integral $\int_{0}^
{\infty} dz\, (\sin z -z) z^{-1-\Lambda}$ that can be calculated by the same
formula (\ref{int2}), i.e.,
\begin{equation}
    \int_{0}^{\infty} dz \frac{\sin z -z} {z^{1+\Lambda}} =
    \frac{\pi}{2\Gamma(1+\Lambda)\cos(\pi\Lambda/2)}.
    \label{int3}
\end{equation}
Thus, the representation (\ref{I4}) remains valid for $\Lambda \in (1,2]$ as
well.

At $\Lambda =1$ both approaches developed for $\Lambda \in (0,1)$ and $\Lambda
\in (1,2]$ are not applicable (the integrals in eqs.~(\ref{int2}) and
(\ref{int3}) are divergent). Therefore, to calculate $I(k)$ at $\Lambda = 1$,
we use the method applied to find $R(k)$ at $\Lambda = 2$. According to that we
write $I(k) = I_{1}(k,\xi) + I_{2}(k,\xi)$, where
\begin{equation}
    I_{1}(k,\xi) = - k \int_{0}^{\xi} dy\, [p(y) - p(-y)]y
    \lim_{\tau \to 0} \frac{a(\tau)}{\tau}
    \label{I(1)}
\end{equation}
and
\begin{equation}
    I_{2}(k,\xi) = - k (h_{+}\delta_{1 \Lambda_{+}}
    - h_{-}\delta_{1 \Lambda_{-}}) \lim_{\tau \to 0} \frac{a
    (\tau)}{\tau}\! \int_{a(\tau)|k|\xi}^{\infty}dz \frac{\sin z}
    {z^{2}}.
    \label{I(2)}
\end{equation}
Then, using the asymptotic formula $\int_{a(\tau)|k|\xi}^{\infty}dz\,
z^{-2}\sin z \sim \ln [1/a(\tau)]$ that occurs as $\tau \to 0$, one obtains
\begin{equation}
    I(k) = - k (h_{+}\delta_{1 \Lambda_{+}}
    - h_{-}\delta_{1 \Lambda_{-}}) \lim_{\tau \to 0} \frac{a
    (\tau)}{\tau} \ln \frac{1}{a(\tau)}.
    \label{I5}
\end{equation}

Next, on the basis of the above derived results we can express the parameters
of the fractional Fokker-Planck equation (\ref{FP1}) through the asymptotic
characteristics of the probability density $p(y)$ (at $y \to \pm \infty$) and
the scale function $a(\tau)$ (at $\tau \to 0$). If $\Lambda \in (0,1)$ or
$\Lambda \in (1,2)$ then, using eqs.~(\ref{R4}) and (\ref{I4}) and the
definition of the Riesz derivative according to which
\begin{eqnarray}
    &\displaystyle \mathcal{F}^{-1} \{|k|^{\Lambda} P_{k}(t) \}
    = - \frac{\partial^{\Lambda}}{\partial |x|^{\Lambda}} P(x,t),&
    \nonumber\\[6pt]
    &\displaystyle \mathcal{F}^{-1} \{ik|k|^{\Lambda-1} P_{k}(t) \}
    = - \frac{\partial}{\partial x}\frac{\partial^{\Lambda-1}}
    {\partial |x|^{\Lambda-1}} P(x,t),&
    \label{rels}
\end{eqnarray}
we obtain that the generalized Fokker-Planck equation (\ref{FP2}) reduces to
the fractional one (\ref{FP1}) with
\begin{eqnarray}
    &\displaystyle \alpha = \Lambda, \quad \beta = \frac{h_{+}\delta_{\Lambda
    \Lambda_{+}} - h_{-}\delta_{\Lambda \Lambda_{-}}} {h_{+}\delta_{\Lambda
    \Lambda_{+}} + h_{-}\delta_{\Lambda \Lambda_{-}}},&
    \nonumber\\[6pt]
    &\displaystyle \gamma = \frac{\pi (h_{+}\delta_{\Lambda \Lambda_{+}}
    + h_{-}\delta_{\Lambda \Lambda_{-}})}{2\Gamma(1 + \Lambda)\sin(\pi
    \Lambda/2)}\,q,&
    \label{abg1}
\end{eqnarray}
and $q = \lim_{\tau \to 0}a^{\Lambda}(\tau)/\tau$. The condition $0<|\ln
S_{k}|<\infty$ assumes that $0<q<\infty$, and so the scale parameter must be
proportional to $\tau^{1/\Lambda}$, i.e., $a(\tau) \propto \tau^{1/\Lambda}$.
We note also that $\beta = (h_{+} - h_{-})/(h_{+} + h_{-})$ if $\Lambda_{+} =
\Lambda_{-} = \Lambda$, $\beta = 1$ if $\Lambda_{-} > \Lambda_{+} = \Lambda$,
and $\beta = -1$ if $\Lambda_{+} > \Lambda_{-} = \Lambda$.

According to eqs.~(\ref{R4}) and (\ref{I5}), at $\Lambda=1$ and $h_{+}\delta_{1
\Lambda_{+}} \neq h_{-}\delta_{1 \Lambda_{-}}$ the condition $0<|\ln S_{k}|<
\infty$ holds only if $a(\tau) \propto \tau/ \ln(1/\tau)$. In this case
$R(k)=0$ and eq.~(\ref{FP2}) takes the form
\begin{equation}
    \frac{\partial}{\partial t}P(x,t) = -\frac{\partial}{\partial x}
    [f(x,t) + f_{0}]P(x,t),
    \label{FP4}
\end{equation}
where
\begin{equation}
    f_{0} = q(h_{+}\delta_{1\Lambda_{+}} - h_{-}\delta_{1 \Lambda_{-}})
    \label{f0}
\end{equation}
and $q = \lim_{\tau \to 0} a(\tau) \ln[1/a(\tau)] /\tau$. Equation (\ref{FP4})
presents a very unexpected and truly remarkable result. It shows that, in
contrast to common belief, a given type of noise \textit{does not} generate a
random dynamics of the system, but rather acts as a \textit{constant} force
$f_{0}$. This force arises from the difference in the asymptotic behavior of
the probability density $p(y)$ at $y \to \infty$ and $y \to -\infty$. If $f_{0}
= 0$, i.e., $h_{+} = h_{-} = h$ and $\Lambda_{+} = \Lambda_{-} = 1$, then
eq.~(\ref{FP4}) becomes invalid, and the next terms of the asymptotic expansion
of $p(y)$ as $y \to \pm \infty$ must be taken into account. For the special
case of symmetric noise, when $p(y) = p(-y)$, eqs.~(\ref{I}) and (\ref{R4})
yield $I(k)=0$ and $R(k)=-|k|\pi h q$, respectively, and thus eq.~(\ref{FP2})
can also be written in the form of eq.~(\ref{FP1}) with
\begin{equation}
    \alpha=1, \quad
    \beta\tan\frac{\pi\alpha}{2} = 0, \quad
    \gamma = \pi h q,
    \label{abg2}
\end{equation}
and $q=\lim_{\tau \to 0} a(\tau)/\tau$, i.e., $a(\tau) \sim q\tau$. We note in
this context that if the same scale function, $a(\tau) \sim q\tau$, is chosen
for asymmetric noise with $h_{+} \neq h_{-}$ then $|I(k)| = \infty$ and,
consequently, the system reaches the final state immediately. This result
clarifies the nature of instabilities in the numerical simulation of $x(t)$
which occur in the present case (see Ref.~\cite{DGH} and references therein).

Finally, at $\Lambda=2$ a non-trivial action of the noise $\xi(t)$ takes place
only if the scale function satisfies the condition $a^{2}(\tau) \sim 2q\tau/
\ln(1/ \tau)$ as $\tau \to 0$. In this case $\lim_{\tau \to 0} a^{2}(\tau)
\ln[1/a (\tau)] /\tau = q$, $\lim_{\tau \to 0} a^{2}(\tau)/\tau = 0$, and, as
it follows from eqs.~(\ref{R5}) and (\ref{I4}), $R(k) = -k^{2}(h_{+}\delta_{2
\Lambda_{+}} + h_{-} \delta_{2 \Lambda_{-}})q/2$ and $I(k)=0$. Therefore, at
$\Lambda = 2$ eq.~(\ref{FP2}) takes the form of the ordinary Fokker-Planck
equation (\ref{FP3}), i.e., eq.~(\ref{FP1}) with
\begin{equation}
    \alpha = 2, \quad
    \gamma = \frac{h_{+}\delta_{2 \Lambda_{+}} +
    h_{-} \delta_{2 \Lambda_{-}}}{2}\,q.
    \label{ag2}
\end{equation}
It should be emphasized that although eq.~(\ref{FP3}) is the same for
$\sigma^{2} \neq \infty$ and $\Lambda=2$, these two cases are quite different
because at $\Lambda=2$ the variance $\sigma^{2}$ of $p(y)$ is infinite. This
difference results in different dependence of the scale parameter $\gamma$ on
$p(y)$. Namely, while in the former case it is proportional to $\sigma^{2}$
[see eq.~(\ref{ag1})], i.e., an \textit{integral} characteristic of $p(y)$, in
the latter case the scale parameter is determined by the \textit{tails} of
$p(y)$ [see eq.~(\ref{ag2})].

Thus, we have shown that (i) each noise whose increments have a heavy-tailed
distribution acts on the system the same (in the sense of the probability
density $P(x,t)$) as a \textit{certain} L\'{e}vy stable noise, and (ii) the
action of each L\'{e}vy stable noise can be reproduced by \textit{different}
noises characterized by different distributions of the noise increments. We
have determined the parameters of the fractional Fokker-Planck equation
(\ref{FP1}) that corresponds to the overdamped Langevin equation
(\ref{Langevin}). If the transition probability density of the noise generating
process is heavy tailed then these parameters are expressed through the
characteristics of the tails. Otherwise, the fractional Fokker-Planck equation
reduces to the ordinary one. These theoretical results seems to be especially
important for the simplification and validation of numerical simulations of the
Langevin systems driven by noises with heavy-tailed distributions of the
increments.

\acknowledgments SID acknowledges the support of the EU through Contract No.
MIF1-CT-2006-021533, and PH acknowledges financial support by the Deutsche
For\-schungs\-ge\-mein\-schaft via the Collaborative Research Centre SFB-486,
Project No. A10, and by the German Excellence Cluster ``Nanosystems Initiative
Munich" (NIM).

\end{document}